\documentclass[mathleft
]{an}

\usepackage{graphicx}
\usepackage{times}

\begin{document}

\title{Front- vs back-illuminated CCD cameras for photometric surveys : a noise budget analysis.}

\author{Crouzet N.\inst{1}\fnmsep\thanks{
  \email{crouzet@obs-nice.fr}\newline} \and Guillot T.\inst{1} \and Fressin F.\inst{1} \and Blazit A.\inst{1} \and the A STEP team}

\titlerunning{title}
\authorrunning{author}
\institute{Observatoire de la C\^{o}te d'Azur, Boulevard de l'Observatoire, BP 4229, 06304 Nice Cedex 4, France}


\abstract{
Exoplanetary transit and stellar oscillation surveys require a very high precision photometry. The instrumental noise has therefore to be minimized. First, we perform a semi-analytical model of different noise sources. We show that the noise due the CCD electrodes can be overcome using a gaussian PSF (Point Spread Function) of full width half maximum larger than 1.6 pixels. We also find that for a PSF size of few pixels, the photometric aperture has to be at least 2.5 times larger than the PSF full width half maximum. Then,  we compare a front- with a back-illuminated CCD through a Monte-Carlo simulation. Both cameras give the same results for a PSF full width half maximum larger than 1.5 pixels. All these simulations are applied to the A STEP (Antarctica Search for Transiting Extrasolar Planets) project. As a result, we choose a front-illuminated camera for A STEP because of its better resolution and lower price, and we will use a PSF larger than 1.6 pixels.
}

\maketitle

\section{Introduction}

The photometric technique allows a direct detection of luminosity variations. Several disciplines are therefore con\-cer\-ned. In asteroseismology, these variations are used to iden\-ti\-fy stellar oscillations, giving an access to  the interior of stars. In planetary sciences, a decrease of luminosity cau\-sed by an extrasolar planet occulting its parent star during a transit is used to characterize the planet (Charbonneau et al. 2000; Moutou et al. 2004). In both cases, a very high precision photometry is required, typically to a milli\-mag\-ni\-tude level. Challenging technical issues have therefore to be solved (Rauer et al. 2004), and high accuracy algorithms are needed (Irwin at al. 2007; Gillon et al. 2006; Magain et al. 2007). The Antarctica Search for Transiting Extrasolar Planets (A STEP) aims to detect planetary transits and stellar oscillations from Dome C, Antarctica (Fressin et al. 2005). The three months continuous night as well as a very dry weather are extremely favorable for photometric surveys. A fully automatized telescope is under development. We present here a first noise analysis of this telescope that leads to the choice of the camera, but that applies to other photometric surveys. The first part shows a noise budget obtained with a semi-analytical model. A second part des\-cribes a Monte Carlo simulation of a front- and a back-illuminated CCD camera.

\section{Noise analysis}

\subsection{Description of the cameras}

When using available commercial cameras and a limited fund, a problem
for photometric surveys is whether to use a backward-illuminated
camera, which maximizes efficiency by having electrodes on the
non-irradiated side of the CCD, or a front-illuminated camera in which
the quantum efficiency is limited by direct reflection on the
electrodes, but which is simpler to build and has thus more pixels for
a si\-mi\-lar price tag. More specifically, in the case of A STEP, these
two classes of cameras were led by:
\begin{itemize}  
\item The back-illuminated camera DW 436 by Andor, with a CCD EEV
  42-40, containing 2048x2048 pixels. The quantuum efficiency peaks at
  94 \%, with a mean of 90 \% in the spectral range 600-800 nm. The pixel size is 13  $\mu$m, and the total CCD size is 2.7
  cm. The same CCD is used for the CoRoT mission, giving us some
  facilities for its testing and characterization. The pixel response
  non-uniformity is around 2 \%.
\item The front-illuminated camera Proline by FingerLake In\-stru\-ments\-,
  with a CCD KAF-16801E by Kodak, containing 4096x4096 pixels. Its
  quantuum efficiency peaks at 63 \%, with a mean of 50 \% in the spectral range 600-800 nm. Around 40 \% of flux is lost with
  respect to the back-illuminated camera. The pixel size is 9 $\mu$m, and the total CCD size is 3.7 cm. The
  pixel response non-uniformity is around 0.5 \%. The front-illuminated ca\-me\-ra has also the advantage to be much cheaper,
  allowing us to purchase a backup one.
\end{itemize}

\subsection{CCD transmission}

\begin{figure}[htbp]
\centering
\resizebox{8cm}{!}{\includegraphics{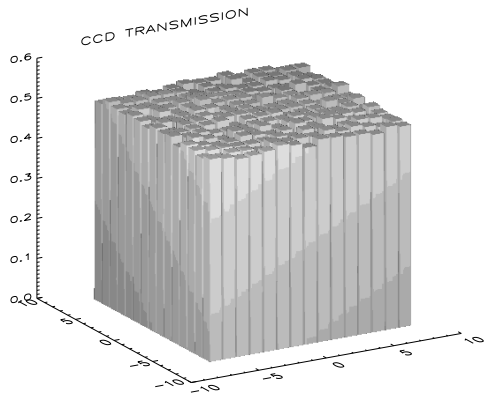}}
\caption{Example of the CCD transmission matrix of a front-illuminated camera. The interpixel variations are typically few per cent. The electrodes cover a part of a pixel and their transmission is very low.}
\label{fig:grille}
\end{figure}

In order to test the CCDs, we model a grid in which the optical transmission of each pixel is
randomly set, with a standard deviation of few per cent. Electrodes
are assumed to cover $\sim 50$ \% of a pixel surface in the case of a
front-illuminated camera, thereby explaining the relatively mo\-dest
quantum efficiency. On the other hand, the back-illu\-mi\-na\-ted camera is
considered as ideal, with no loss due to the electrodes. An example of a
CCD transmission is shown in figure~\ref{fig:grille}.


\subsection{The sources of instrumental noise}
We perform a first analysis of the different noise
sources. A good understanding of noises is indeed necessary for the choice of a camera. In a more general way, this is critical in transit survey data analysis (Pont, Zucker \& Queloz 2006; Smith et al. 2006). All noises are calculated in a semi-a\-na\-ly\-ti\-cal model
using squared photometric apertures of 3x3 and 5x5 pixels. We use
the A STEP instrumental characteristics. The telescope is a 40 cm with
a F/D of 4. Stars range from magnitude 11 to magnitude 16. The exposure time is 10 seconds. The PSF (Point Spread Function) is a two-dimensional gaussian
function. These parameters are preliminar and will have to be
defined more precisely during the telescope design phase, in order to optimize the survey (Horne 2002). Noises
correspond to one exposure, without any image processing. This is
therefore a worst case. We consider the following noise sources :

\begin{itemize}  

\item Electrode noise : The bad optical
transmission of electrodes leads to a loss of flux. This loss depends
on the PSF position with respect to the electrodes. The PSF motion
onto the CCD, due to the telescope jitter, leads to loss
variations. The resulting noise is therefore calculated as the
variation of the flux hitting the electrodes.  
\\

\item Overflow noise : Because of light sources such as crow\-ding
in the field of view, sky brightness, etc., a photometric aperture is
set around each target star. The flux outside this aperture is eliminated,
shoud it come from the target star or another source. This results in
a loss of the flux from the star if the PSF overflows the photo\-me\-tric
aperture. The loss depends again on the PSF position inside this
aperture, and varies due to the telescope jitter. The
noise is then calculated as the variation of the flux inside the
aperture.
\\

\item Interpixel noise : Each pixel has its own optical
transmission, which vary from one pixel to another by typically 1 per
cent. This PRNU (Photo Response Non Uniformity) is
taken into account defining an equi\-valent number $ N $ of pixels under
the PSF. The resulting noise is :

\begin{eqnarray*}
\sigma_{interpx}=\frac{PRNU}{\sqrt{N}}
\end{eqnarray*}

were $ PRNU $ is the standard deviation of the pixel transmission distribution.
\\

\item Other noise sources : Other noise sources are implemented such as the photon noise from the target star, the noise from the sky background (taken as 22  mag/arcsec$^2$ with a slow variation along the CCD), and the camera dark current and read-out noise.

\end{itemize}  


\begin{figure*}[htbp]
\centering
\resizebox{11cm}{!}{\includegraphics{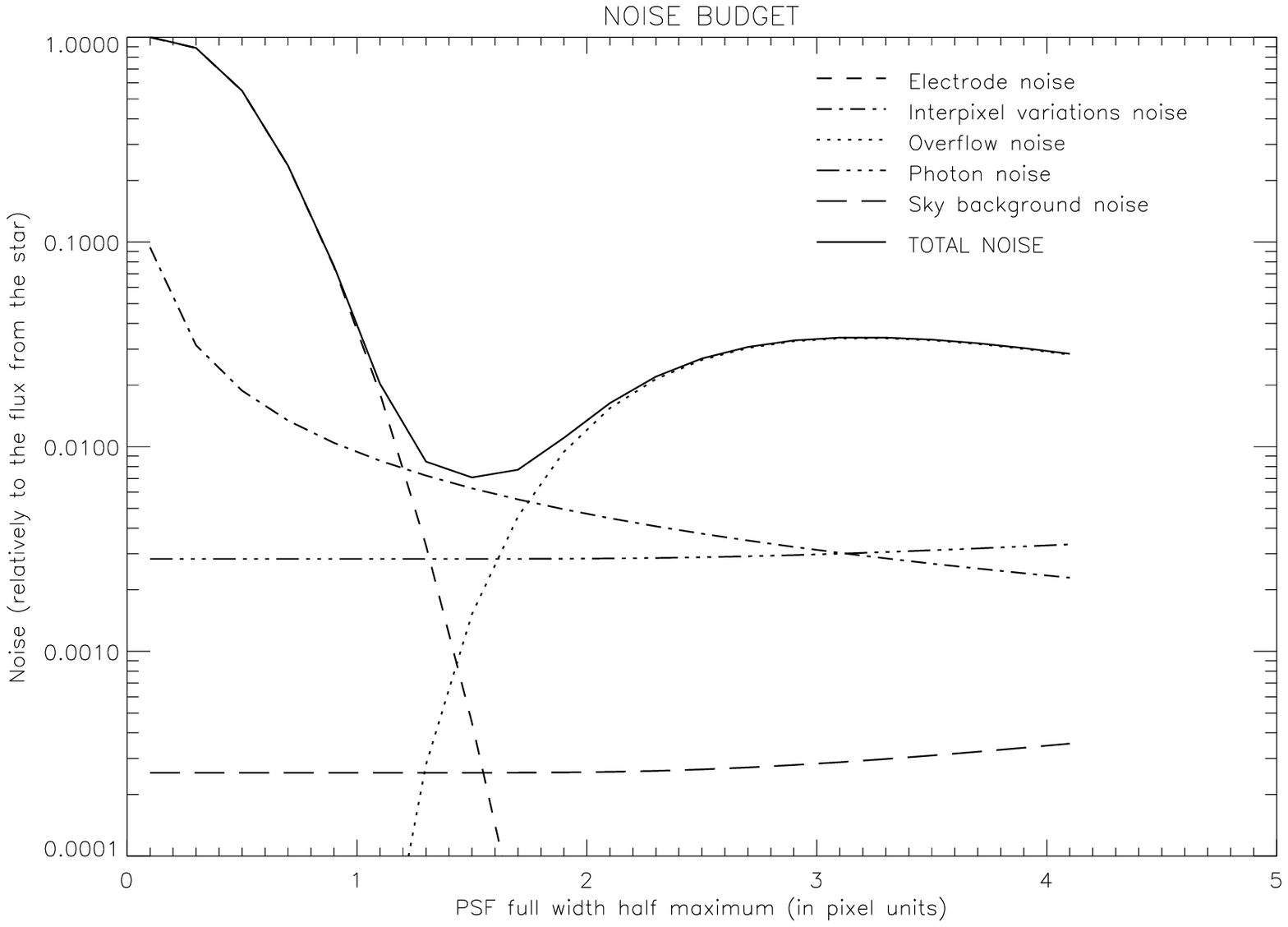}}
\resizebox{11cm}{!}{\includegraphics{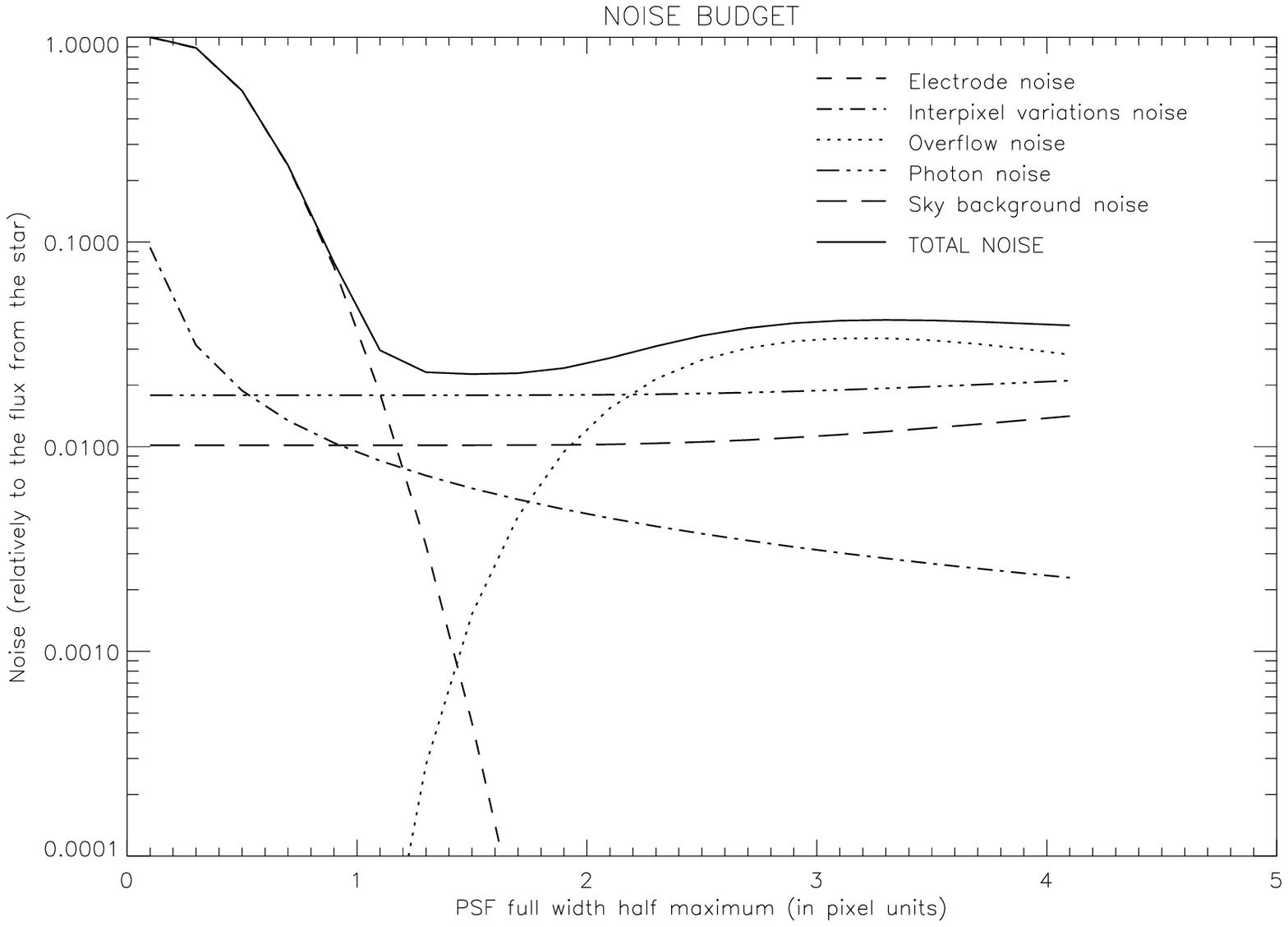}}
\caption{Noise budget as a function of the PSF full width half maximum, in a semi-analytical model, for a front illuminated camera with a 5x5 pixels photometric aperture and 1 \% interpixel variations. The results for stars of magnitude 11 (top) and 15 (bottom) are represented. The CCD dark current and read-out noise are not plotted, and are always lower than the sky background noise.
}
\label{fig:noises_budget}
\end{figure*}

\subsection{A semi-analytical model}
The results of the analysis based on the semi-analytical mo\-del are
represented in figure~\ref{fig:noises_budget}. The electrode and
overflow noises are clearly dominant. The interpixel noise is also
do\-minant for bright stars. As supposed, the other noise sources are not dominant for a 11 magnitude star. For a 16 magnitude star, the photon noise is dominant for full width half maxima between 1.2 and 2 pixels, but does not change the total noise curve shape.

The main observation is that the electode
noise reaches a level of $10^{-4}$ for PSF full width half
maxima larger than 1.6 pixels. This noise can therefore be overcome adjusting the PSF size. (It can be noted however, that the noise
strongly increases when considering non-gaussian PSFs with sharp
interfaces, e.g. a top-hat function. We will not consider this
further.) We also see that the overflow noi\-se becomes do\-mi\-nant for full width half maxima larger than 2 pixels, given the aperture of
5x5 pixels we use. In a general way, for our PSFs of few pixels, we
find that the photometric aperture must be at least 2.5 larger than
the PSF full width half ma\-xi\-mum.

\subsection{Monte-Carlo simulations}

\begin{figure*}[htbp]
\centering
\resizebox{11cm}{!}{\includegraphics[angle=0]{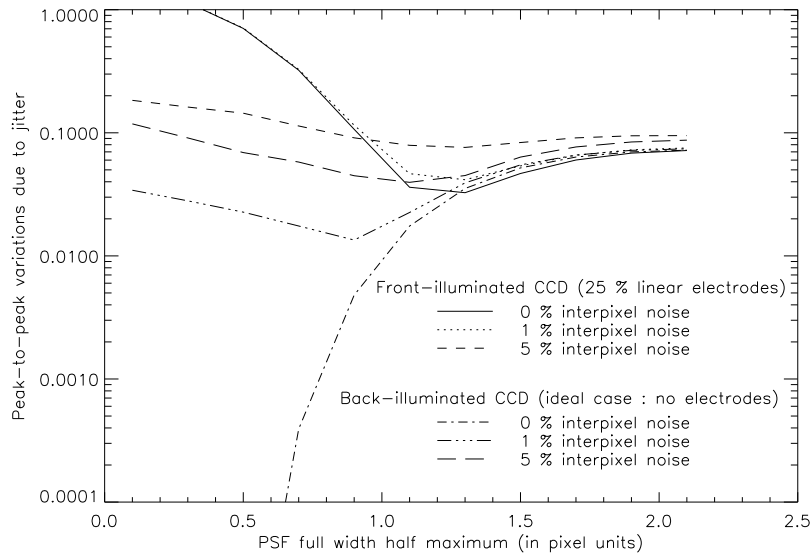}}	
\caption{Noise as a function of the PSF full width half maximum, for a 3x3 pixels photometric aperture. A front illuminated camera (with electrodes), and a back-illuminated camera (as an ideal case with no electrodes) are represented. Three
values of interpixel variations are used.}
\label{fig:simcam_summary}
\end{figure*}

In order to test our two cameras, we use a direct simulation of the
CCD that attempts to mimic real observations in\-clu\-ding jitter and
interpixel noise. The simulations proceed as follows : during a run, a gaussian PSF is moved along the CCD in an arbitrary
direction. For each position, the flux inside the photometric aperture
is measured. The resulting noise level is estimated as the peak-to-peak flux variation. This noise is calculated for several PSF full
width half ma\-xi\-ma, since this parameter can be chosen during the
te\-les\-cope design phase. This simulation is performed for three
interpixel values, 0, 1, and 5 \% as the real value is unknown, and for
two electrode sizes, 50 and 0 \% of a pixel, i.e. for both cameras.  The
results of the Monte-Carlo simulations are shown in figure~\ref{fig:simcam_summary}. Note this simulation does not account
for different quantuum efficiencies of both cameras. The advantage
of the back-illuminated camera, i.e. no electrodes, is valid
only for PSFs with full width half maxima smaller than 1.5 pixels. For
larger PSFs, both cameras give the same results, which means that the electrode
noise is negligible. This is in perfect agreement with our
semi-analytical model.


These simulations imply that for well-sampled PSFs, the noise
difference is essentially due to the difference in quantuum efficiency,
i.e. for the two cameras that are considered,
$\sqrt{0.9/0.5}=1.34$. This is to be compared to the fact that the 4
times increase in pixel number allows having either 4 times as many targets
with the same crowding, or, with exactly the same field of view, a
reduction of the interpixel noise by a factor 2 due to PSFs that are
better sampled spatially. 

Other advantages of the presently built CCD cameras is that
front-illuminated ones have less interpixel noise (1\% vs. 3\%), and
are generally cheaper (by $\sim 30$\%). A front-illuminated CCD camera
is therefore, in the case of A STEP, more advantageous.

\section{Conclusion}

We performed a semi-analytical noise analysis to identify the limiting
noise sources in precision photometry, using a gaussian PSF. The
electrode, overflow, and interpixel noises are dominant, as well as
the photon noise for faint stars. We showed that the electrode
noise becomes negligible for gaussian PSF full width half maxima
larger than 1.6 pixels. We also found that for PSFs of few pixels,
the photometric aperture must be at least 2.5 times larger than the
PSF full width half maximum. 

We then compared a front- and
a back-illuminated ca\-me\-ra in a Monte Carlo simulation. For
photometric surveys for which the PSF is well-sampled (at least 1.5
pixels full width half maximum), and limited in terms of budget to exis\-ting commercial
cameras, we found that a front-illu\-mi\-na\-ted camera is a better
alternative.

{}

\end{document}